\documentclass[twocolumn,showpacs,preprintnumbers,amsmath,amssymb]{revtex4}
\usepackage{graphicx}
\usepackage{bm}
\usepackage{dcolumn}

\def\prl#1#2#3{{ Phys.   Rev.   Lett.  } {\bf #1}, #2 (#3)}

\def\pre#1#2#3{Phys.   Rev.   E {\bf #1}, #2 (#3)}
\def\prb#1#2#3{Phys.   Rev.   B {\bf #1}, #2 (#3)}
\def\prlon#1#2#3{Proc.   Phys.   Soc.   London A {\bf #1}, #2 (#3)}
\def\pra#1#2#3{Phys.   Rev.   A {\bf #1}, #2 (#3)}

\def\jpa#1#2#3{J.   Phys.   A {\bf #1}, #2 (#3)}

\def\rmp#1#2#3{Rev.   Mod.   Phys.   {\bf #1}, #2 (#3)}
\def\ijotp#1#2#3{Int. J. of Thermo Phys. {\bf#1},#2 (#3)}
\def\zpb#1#2#3{Z Phys. B {\bf#1},#2 (#3)}
\def\ssin#1#2#3{Sci. Sin. {\bf#1},#2 (#3)}
\def\prd#1#2#3{Phys. Rev. D {\bf#1},#2 (#3)}
\def\prep#1#2#3{ Phys. Rep. {\bf#1},#2 (#3)}

\def\noi{\noindent}
\def\bc{\begin{center}}
\def\ec{\end{center}}
 \newcommand{\bea}{\begin{equation}}
 \newcommand{\eea}{\end{equation}\noi}
 \newcommand{\ber}{\begin{eqnarray}}
 \newcommand{\eer}{\end{eqnarray}\noi}
\begin{document}
\title{Non equilibrium statistical physics with a fictitious time}
\author{Himadri    S.   Samanta}\email{tphss@mahendra.iacs.res.in}
\author{J.   K. Bhattacharjee}\email{tpjkb@mahendra.iacs.res.in}
\affiliation{Department of Theoretical Physics,
Indian Association for the Cultivation of Science \\
Jadavpur, Calcutta 700 032, India}
\date{\today}
\begin{abstract}
Problems in non equilibrium statistical physics are 
characterized by the absence of a fluctuation dissipation theorem.
The usual analytic route for treating these vast class of problems
is to use response fields in addition to the real fields that are 
pertinent to a given problem. This line of argument was introduced
by Martin, Siggia and Rose. We show that instead of using the response 
field, one can, following the stochastic quantization of Parisi and Wu, 
introduce a fictitious time. In this extra dimension a fluctuation
dissipation theorem is built in and provides a different outlook to 
problems in non equilibrium statistical physics.

\end{abstract}

\newpage

\maketitle

The fluctuation dissipation theorem (FDT) is the central feature
in the study of dynamics of statistical systems, when the deviation
from equilibrium is small. The response of such a system to a small external
probe is studied within a linear response theory.This is a macroscopic effect
which can be observed in a dynamic measurement. The FDT relates 
this response to the fluctuation properties of the thermodynamic system.
In its symplest form it appears in Brownian motion-the molecular fluctuation
responsible for the agitation of the suspended particles are the same as those
responsible for the macroscopic friction experienced by the particles in 
motion. Generalised, it says that in an equilibrated system under a small 
perturbation there is a balance between a systematic force and a 
chaotic fluctuating force. The theorem is valid for all dynamical 
systems in which an equilibrium distribution is reached at infinitely 
long times. Clearly, for a system which is far from equilibrium this 
theorem is absent. In this communication, we would like to point out
that following the technique of stochastic quantization, we can introduce 
an extra time like dimension in these far from equilibrium problems 
and in this fictitious time introduce a FDT. This greatly simplifies
 the study of possible scaling solutions in the problem.   

The FDT facilitates the study of dynamics near equilibrium by the existence 
of the fluctuation dissipation theorem which relates the correlation 
function to the response function. One of the primary difficulties in 
the study of dynamics far from equilibrium is the absence of a 
fluctuation dissipation theorem leading to independent diagrammatic 
expansions for the response function and the correlation function. 
The difficulty stems from the fact that the equilibrium distribution
is not known and the only way in which we can do an averaging is the averaging
over noise. The spectacular success in the last few decades in the 
study of dynamics has been in the near equilibrium dynamics near
second order phase transitions. Success in this case implies
a close and detailed correspondence between theory and
experiment in a variety of systems \cite{1,2,3,4,5,6}.
The dynamics in this case has typically 
been that of a n-component field $\phi_{\alpha}(\alpha =1,2,....,n)$ 
satisfying an equation of motion (model $A$, model $B$ etc.)

\bea\label{eq1}
\frac{\partial \phi_{\alpha}}{\partial t}= -\Gamma 
\frac{\delta F}{\delta \phi_{\alpha}}+\eta_{\alpha}
\eea

where $ F $ is a free energy functional (most often the Ginzburg-Landau 
free energy), with the Gaussian white noise $\eta(\vec{r},t)$ 
satisfying

\bea\label{eq2}
<\eta_{\alpha}(\vec{r_{1},t_{1}})\eta_{\beta}(\vec{r_{2}},t_{2})>
=2\Gamma kT \delta_{\alpha \beta} 
\delta(\vec{r_{1}}-\vec{r_{2}})\delta(t_{1}-t_{2})
\eea

It is straightforward to check by going to the associated 
Fokker Planck equation that $ exp(-F/kT)$ is indeed the 
equilibrium distribution. Since the dynamics is always close to equilibrium,
it is the averaging with the distribution $ exp(-F/kT)$ which
one has in mind. There have been more complicated equations of 
motion with the structure

\bea\label{eq3}
\frac{\partial \phi_{\alpha}}{\partial t}=
V_{\alpha}([\phi_{\beta}])-\Gamma 
\frac{\delta F}{\delta \phi_{\alpha}}+\eta_{\alpha}
\eea

(e.g. models $E,F,G,H,J$ of dynamic critical phenomena), but 
the equilibrium distribution has been maintained because
$\frac{\partial \phi_{\alpha}}{\partial t}=
V_{\alpha}([\phi_{\beta}])$ keeps the free energy $F$ constant in time
(in fact that is the motivation behind the construction of $V_{\alpha}$).

Models of non equilibrium statistical mechanics do not evolve towards
an equilibrium distribution. Consequently, the averaging involved
in the construction of correlation function have to be done
explicitly over the noise inherent in the statistical dynamics 
(the noise written explicitly in Eqs (\ref{eq1}) and (\ref{eq2})). 
One of the most-studied models in this category is the Kardar 
Parisi Zhang (KPZ) equation, where the dynamics ( physically the growth
of a surface by deposition of atoms on a substrate ) is given by

\bea\label{eq4}
\frac{\partial \phi(\vec{r},t)}{\partial t}=
\nu \nabla^{2}\phi - \frac{\lambda}{2} 
(\vec{\nabla}\phi)^{2}+ \eta
\eea

where
\bea\label{eq5}
<\eta(\vec{r_{1}},t_{1})\eta(\vec{r_{2}},t_{2})>=
2D_{0} \delta(\vec{r_{1}}-\vec{r_{2}})\delta(t_{1}-t_{2})
\eea

The KPZ equation models a growing interface which develops 
from a random deposition of atoms on a substrate of dimensions $D$.
The randomness in the deposition process brings in the noise term
in the time evolution of the height of the interface. The deterministic part
of the evolution coming from the deposition and evaporation of atoms
is governed by the chemical potential $\mu$ which is determined by
the total height gradient and its derivatives. Clearly a linear term 
in the gradient is not acceptable since the evolution cannot depend 
of the sign of the gradient. Thus the linear term in $\mu$
has to be the second derivative of the height and the 
non linear term is the square of the gradient. Without the non linear
term, one has the Edwards Wilkinson model \cite{8} of growth and with 
it the KPZ model. One of the primary uses in the models of growth
is the question about the roughness of the interface. 
This is expressed in terms of the correlation function 
$<\phi(\vec{x}+\vec{r},t)\phi(\vec{x},t)>$, which scales as $r^{2\alpha}$
for large $r$. If $\alpha >0$,then the height fluctuations are coupled at long 
distances at the surface is rough. If $\alpha <0$, the fluctuations 
decay and the surface is smooth. The EW model has $\alpha=(2-D)/2$ 
leading to a rough interface for $D \leq 2$. For the KPZ model, $\alpha >0$
and the surface is always rough for $D<D_{c}$, an upper critical dimension,
while for $D>D_{c}$, the surface is smooth for small values of 
the coupling constant. The value of $D_{c}$ is still controversial
\cite{9,10,11,12}.

Now Eq. (\ref{eq4}) is not of the form of Eq.(\ref{eq1})
 ($\nu \nabla^{2}\phi - \frac{\lambda}{2}(\vec{\nabla}\phi)^{2}$
cannot be written in the form $\delta F/\delta \phi$). It is of the 
form of Eq. (\ref{eq3}), with 
$F=(\nu/2)\int{d^{D}x(\vec{\nabla}\phi)^{2}}$, but identifying
$V([\phi])$ as $-(\lambda/2)(\vec{\nabla}\phi)^{2}$, it can be shown that
$F$ is preserved only in $D=1$ ($D$ is the substrate dimension).  

The field theoretic technique of handling such problems was 
first explained by Martin, Siggia and Rose \cite{13} and perfected over 
the years by 
Janssen \cite{14}, Bausch et al \cite{15}, 
De Dominicis and Peliti \cite{16},
 Frey and T\"{a}uber \cite{17}, Plischke et al \cite{18}
and a host of other researchers. 
One begins by writing the "partition function" 

\ber\label{eq6}
& & Z=\int\mathcal{D}\phi\int\mathcal{D}\eta
e^{-\frac{1}{2D_{0}}\int d^{D}x dt \eta(\vec{r},t)^{2}}\nonumber\\& &
\delta(\frac{\partial \phi}{\partial t}- \nu \nabla^{2}\phi
-\frac{\lambda}{2}(\vec{\nabla}\phi)^{2}-\eta(\vec{r},t))\nonumber\\& &
\eer

Integrating over $\eta$, $Z=\int \mathcal{D}\phi exp-S(\phi)$,
with the action

\bea\label{eq7}
S(\phi)=\frac{1}{D_{0}}\int d^{D}\vec{r} dt 
[\frac{\partial \phi}{\partial t}-\nu \nabla^{2}\phi
-\frac{\lambda}{2}(\vec{\nabla}\phi)^{2}]^{2}
\eea
as given by Zee.

It is with the action that all correlations of the form 
$<\phi(\vec{r_{1}},t_{1})\phi(\vec{r_{2}},t_{2})>$ have to be 
determined. The response functions are introduced by writing
$Z$ in terms of an auxilliary field, so that 

\bea\label{eq8}
Z=\int \mathcal{D}[\phi]\mathcal{D}[\tilde{\phi}]
e^{-S(\phi ,\tilde{\phi} )}
\eea
where the new action is given by

\bea\label{eq9}
S(\phi ,\tilde{\phi})=\int dt \int d^{D}\vec{r}
[\tilde{\phi}(\vec{r},t)[\dot{\phi}-
\nu \nabla^{2}\phi-\frac{\lambda}{2}(\vec{\nabla}\phi)^{2}]-D\tilde{\phi}^{2}]
\eea

The way of handling this case of non equilibrium
statistical mechanics delineated clearly in Frey and T\"{a}uber \cite{17}. 
Our proposal, here, is to use the action of (\ref{eq7}) and 
exploit the principle of stochastic quantization to set 
up an alternative approach to the study of this class of problems. 
Stochastic quantization is a method of quantisation 
proposed by Parisi and Wu \cite{19,20} based on stochastic 
Langevin dynamics of a physical
system in a fifth time $\tau$. They showed that at the perturbative level,
the usual quantum field theory was recovered in the limit 
$\tau \rightarrow \infty$ of this dynamics. Euclidean quantum field 
theory correlation functions for a field $\phi$ corresponding 
to an action $S(\phi)$ are given by

\ber\label{eq10}
& & <0\mid T \phi (x_{1})\phi (x_{2})......\phi (x_{l})\mid 0>\nonumber\\
&=&\frac{\int \mathcal{D}[\phi]\phi(x_{1})....\phi(x_{l})e^{-S[\phi]}}
{\int \mathcal{D}[\phi]e^{-S[\phi]}}\nonumber\\& &
\eer

Parisi and Wu proposed the following alternative method:
(a)  Introduce an extra fictitious "time" $\tau$ in addition to the four 
space time $X^{\mu}$ and postulate a Langevin dynamics

\bea\label{eq11}
\frac{\partial \phi(x,t,\tau)}{\partial \tau}=
-\frac{\delta S}{\delta \phi}+f(x,t,\tau)
\eea
where $f$ is a Gaussian random variable with
\bea\label{eq12}
<f(x,t,\tau)f(x^{\prime},t^{\prime}\tau^{\prime})>=
2\delta(x-x^{\prime})\delta(t-t^{\prime})\delta(\tau-\tau^{\prime})
\eea

(b)  Evaluate the stochastic average of the fields
$\phi(x,t,\tau)$ satisfying Eq.(\ref{eq11}) ie evaluate
$<\phi(x_{1},\tau_{1})\phi(x_{2},\tau_{2})...\phi(x_{l},\tau_{l})>_{\eta}$

(c)  Set $\tau_{1}=\tau_{2}=....=\tau_{l}=\tau$ and take the limit 
$\tau \rightarrow \infty$. Then one has 
\ber\label{eq13}
& &\lim_{\tau \to\infty}<\phi(x_{1},\tau)\phi(x_{2},\tau)
....\phi(x_{l},\tau)>_{\eta}\nonumber\\
&=&\frac{\int \mathcal{D}[\phi]\phi(x_{1})\phi(x_{2})....\phi(x_{l})
e^{-S[\phi]}}{\int \mathcal{D}[\phi]e^{-S[\phi]}}\nonumber\\& &
\eer

We now return to the KPZ equation and write the action of Eq.(\ref{eq7})
in momentum-frequency space as

\ber\label{eq14}
& & S=\frac{1}{4D_{0}}\int \frac{d^{D}k}{(2\pi)^{D}}
\frac{dw}{2\pi}\{(w^{2}+\nu^{2}k^{4})\phi(\vec{k},w)\phi(-\vec{k},-w)
\nonumber\\
& & + \frac{\lambda}{2}\sum_{\vec{q},w^{\prime}}(-iw+\nu k^{2})
\vec{q}.(\vec{k}+\vec{q})
\phi(\vec{k},w)\phi(\vec{q},w^{\prime}) \nonumber\\ & &
\phi(-\vec{k}-\vec{q},-w-w^{\prime})
-\frac{\lambda}{2}\sum_{\vec{q},w^{\prime}}(iw+\nu k^{2})
\vec{q}.(\vec{k}-\vec{p})\nonumber\\& &
\phi(\vec{q},w^{\prime})
\phi(-\vec{k}-w)\phi(\vec{k}-\vec{q},w-w^{\prime})\nonumber\\& &
-\frac{\lambda^{2}}{4}\sum_{\vec{p},\vec{q},w^{\prime},w^{\prime\prime}}
[\vec{p}.(\vec{k}-\vec{p})][\vec{q}.(\vec{k}+\vec{q})]\nonumber\\& &
\phi(\vec{p},w^{\prime})\phi(\vec{k}-\vec{p},w-w^{\prime})
\phi(-\vec{k}-\vec{q},-w-w^{\prime\prime})\phi(\vec{q},w^{\prime\prime})\}
\nonumber\\
&&
\eer
in accordance with standard results \cite{21}.
The Langevin equation in the fifth time $\tau$ is now written as

\bea\label{eq15}
\frac{\partial \phi(\vec{l},\Omega,\tau)}{\partial \tau}=
\frac{\delta S}{\delta \phi(-\vec{l},-\Omega,\tau)}+f
\eea
with

\bea\label{eq16}
<f(\vec{l},\Omega,\tau)f(\vec{l}^{\prime},\Omega^{\prime},\tau^{\prime})>
=2\delta(\vec{l}+\vec{l}^{\prime},\Omega+\Omega^{\prime}
,\tau-\tau^{\prime})
\eea

After straightforward algebra Eq.(\ref{eq15}) acquires the form
\ber\label{eq17}
& &\frac{\partial \phi(l,\Omega,\tau)}{\partial \tau}=
-\frac{(\Omega^{2}+\nu^{2}l^{4})}{2D_{0}}\phi(l,\Omega,\tau)\nonumber\\
&-&\frac{\lambda}{4D_{0}}\sum_{\vec{k},w}[2(-iw+\nu k^{2})
\vec{l}\dot (\vec{l}-\vec{k})\nonumber\\
&-&(i\Omega+\nu l^{2})
\vec{k} . (\vec{l}-\vec{k})]\phi(\vec{k},w,\tau)
\phi(\vec{l}-\vec{k},\Omega-w,\tau )\nonumber\\&+&
\frac{\lambda^{2}}{4D_{0}}\sum_{\vec{k},\vec{q},w,w^{\prime}}
[\vec{l}. (\vec{l}-\vec{k})][\vec{q}. (\vec{k}-\vec{q})]
\phi(\vec{q},w^{\prime},\tau)\nonumber\\
& &\phi(\vec{k}-\vec{q},w-w^{\prime},\tau^{\prime})
\phi(\vec{l}-\vec{k},\Omega-w,\tau^{\prime})
+f(l,\Omega,\tau)\nonumber\\
& &
\eer

For the free field theory, (Edwards Wilkinson model)
$\lambda=0$ and we get the response function of the system as
\bea\label{eq18}
G_{0}=(-iw_{\tau}+\frac{w^{2}+\nu^{2}k^{4}}{2D_{0}})^{-1}
\eea
where $w_{\tau}$ is the Fourier transform variable
corresponding to the fictitious time $\tau$. The correlation function is

\bea\label{eq19}
C_{0}(k,w,w_{\tau})=[w_{\tau}^{2}+\frac{(w^{2}+\nu^{2}k^{4})^{2}}
{4D_{0}^{2}}]^{-1}=\frac{1}{w_{\tau}}Im G_{0}
\eea
as required by the fluctuation dissipation theorem
in the fictitious time. To get the correlation function 
of the original theory, we need to consider $C(k,w,\tau_{1},\tau_{2})$; 
set $\tau_{1}=\tau_{2}=\tau $ and let $\tau \rightarrow \infty$.
The part which survives when $\tau \rightarrow \infty$, is obtained 
directly from the equal-$\tau$ from Eq.(\ref{eq19}), ie. the result which is
obtained by integrating the right hand side of Eq.(\ref{eq19}) over $w_{\tau}$.
This yields $2D_{0}(w^{2}+\nu^{2}k^{4})^{-1}$ for the 
correlation function of the Edwards Wilkinson model, leading to the result
$\alpha=(2-D)/2$.

With $\lambda \not= 0$, we develop as usual the fully dressed 
Greens function from perturbation theory and write

\bea\label{eq20}
G=-iw_{\tau}+\frac{w^{2}+\nu^{2}k^{4}}{D_{0}}+\Sigma(k,w,w_{\tau})
\eea
The point of our approach is that the $w=0$ part of $\Sigma$ 
gives the flow $\nu$ and the $k=0$ part of $\Sigma$ gives the flow of $D$.
Further, mode coupling analysis will directly yield the exponent $z$
from a power counting analysis. As in other studies the result 
$ \alpha + z=2$ holds in this case as well.

If we inserted in $\Sigma(k,w,w_{\tau})$ to $O(\lambda^{2})$,
then we note that the contribution comes from two different sources - a one
loop contribution from the $O(\lambda^{2})$ term in Eq.(\ref{eq17})
and a two loop contribution from the $O(\lambda)$ term. The
contribution of the $O(\lambda^{2})$ term can be read off from Eq.(\ref{eq17})
by contracting two of the $\phi$-fields. The resulting correction to 
$ (\Omega^{2}+\nu^{2}l^{4})/2D_{0}$ does not have any new momentum
dependence and hence it is the second order contribution from the 
$O(\lambda)$ term which is of significance. The expression for
$\Sigma(l,\Omega,w_{\tau})$ is found after standard alegbra as

\ber\label{eq21}
& &\Sigma(l,\Omega,w_{\tau})=
\frac{\lambda^{2}}{8D_{0}^{2}}
\int \frac{d^{D}k}{(2\pi)^{D}}\frac{dw}{2\pi}\frac{dw_{\tau}}{2\pi}
\nonumber\\& &
\{ [ 2(-iw+\nu k^{2})\vec{l}\cdot (\vec{l}-\vec{k})
-(-\Omega +\nu l^{2})\vec{k}\cdot (\vec{l}-\vec{k})]\nonumber\\ & &
[2(-iw +i\Omega+\nu(\vec{l}-\vec{k})^{2}\vec{l}\cdot \vec{k})
+(iw+\nu k^{2})\vec{l}\cdot (\vec{l}-\vec{k})]\nonumber\\ & &
G(k,w,w_{\tau}^{\prime})C(\vec{l}-\vec{k},\Omega-w,w_{\tau}-w_{\tau}^{\prime})
\nonumber\\ & &+
similar \ term \ interchange \ of\ \vec{k}\ and\ (\vec{l}-\vec{k})
\nonumber\\& & 
except\ in\ the\ first\ factor \}\nonumber\\& &
\eer

We now note that the Greens' function can be written to $O(\lambda^{2})$
as
\bea\label{eq22}
G^{-1}(l,w,w_{\tau})=-iw_{\tau}+\frac{1}{2D_{0}}
(\Omega^{2}+\nu^{2}_{eff}l^{4})
\eea
where
\bea\label{eq23}
\nu^{2}_{eff}l^{4}=\nu^{2}l^{4}+
\Sigma(l,\Omega,w_{\tau})
\eea
or

\bea\label{eq24}
\nu_{eff}l^{2}\simeq \nu l^{2}+
\frac{1}{2\nu l^{2}}\Sigma(l,\Omega,w_{\tau})
\eea
or
\bea\label{eq24a}
\Delta \nu l^{2}=\frac{1}{2\nu l^{2}}\Sigma(l,\Omega,w_{\tau})
\eea

In a self consistent mode coupling, we now replace
$\nu$ by $\Delta \nu $ in Eq.(\ref{eq21}), use $G$ as given 
by Eq.(\ref{eq22}) and $C$ as follows from the 
fluctuation dissipation theorem. We can carry out
the momentum count of Eq.(\ref{eq21}),keeping in mind that 
$\nu l^{2}\sim l^{z}$, to find

\bea\label{eq25}
\Sigma \sim l^{D}l^{z}l^{2z+4}l^{-4z}=l^{D+4-z}
\eea
Using this in Eq.(\ref{eq24a}), we have
$l^{z}\sim l^{D+2-z}$, leading to 
\bea\label{eq26}
z=1+\frac{D}{2}
\eea

Our analysis is not valid for $D>2$, because then $\Sigma$
no longer dominates $l^{2}$. We have checked to that part from 
Eq.(\ref{eq21}) and Eq.(\ref{eq25}), one can write down flows for
$\nu $ $D_{0}$ and as in the usual analysis, there is a critical
$D$, $1<D<2$, where the flow of the coupling constant 
$\lambda^{2}D_{0}/\nu^{3}$ changes sign. We do not consider
that to be of any particular significance one way or another. 
Our primary claim is that by introducing a fictitious time,
we are able to bring back to fluctuation dissipation theorem and 
this allows a power counting scaling analysis for the 
KPZ system which had not seen possible before. 

In closing, we would like to reiterate that the standard practice
in dealing with out of equilibrium situations has been to use a 
set of fictitious fields called response fields, which provides 
a field theoretic prescription for the response function.
Instead, we have proposed the introduction of a fictitious time
in which a FDT holds and thereby only correlation functions need 
to be calculated. Study of scaling solutions is thereby greatly
simplified.


\begin{thebibliography}{99}

\bibitem{1} P. C. Hhenberg and B. I. Halperin, \rmp{49}{435}{1977}
\bibitem{2} A. Onuki, "Phase transition dynamics", Cambridge Univ. Press,
Cambridge (2002)
\bibitem{3} J. V. Sengers, \ijotp{6}{203}{1985}
\bibitem{4} For the superfluid transition see, G. Ahlers, P. C. Hohenberg 
and A. Kornblit, \prl{46}{493}{1981}, V.Dohm and R. Folk, \prl{46}{349}{1981}
\bibitem{5} For the classical fluid, see, R. F. Berg, M. R. Moldover and 
G. A. Zimmerli, \prl{82}{920}{1999}, \pre{60}{4079}{1999}, J. K. Bhattacharjee 
and R. A. Ferrell, \pra{27}{1544}{1983}, H. Hao, R. A. Ferrell and 
J. K. Bhattacharjee, \pre{71}{021201}{2005}
\bibitem{6} For the magnetic transition, see, L. Passell,
O. W. Dietrich and J. Als-Nielsen, \prb{14}{4897}{1976}, J. K\"{o}tzler 
and H. V. Philipsborn, \prl{40}{790}{1978}, J. K. Bhattacharjee 
and R. A. Ferrell, \prb{24}{6480}{1981}, J. K. Bhattacharjee, 
\prb{27}{3058}{1983}, E. Frey and F. Schwabl, \zpb{71}{355}{1988}
\bibitem{7} M. Kardar, G. Parisi and Y. C. Zhang, \prl{56}{889}{1986}, 
see also T. Halpin-Healy and Y. C. Zhang, \prep{254}{215}{1995}
\bibitem{8} S. F. Edwards and D. R. Wilkinson, \prlon{381}{17}{1982}
\bibitem{9} M. L\"{a}ssig and H. Kinzelbach, \prl{78}{903}{1997}
\bibitem{10} C. Castellano, M. Marsili and L. Pietronero, \prl{80}{3527}{1998}
\bibitem{11} J. K. Bhattacharjee, \jpa{31}{L93}{1998}
\bibitem{12} E. Marinari, A. Pagnani and G. Parisi, \jpa{33}{8181}{2000}
\bibitem{13} P. C. Martin, E. D. Siggia and H. A. Rose, \pra{8}{423}{1973}
\bibitem{14} H. K. Janssen, \zpb{23}{377}{1976}
\bibitem{15} R. Bausch, H. K. Janssen and H. Wagner, \zpb{24}{113}{1976}
\bibitem{16} C. De Dominicis and L. Peliti, \prb{18}{353}{1978}
\bibitem{17} E. Frey and U. C. T\"{a}uber, \pre{50}{1024}{1994}
\bibitem{18} T. Sun and M. Plischke, \pre{49}{5046}{1994}
\bibitem{19} G. Parisi and Y. S. Wu, \ssin{24}{484}{1981}
\bibitem{20} E. Gozzi, \prd{28}{1922}{1978}
\bibitem{21} A. Zee: "Quantum Field Theory in a Nutshell", Universities Press 
(2005)     




\end{thebibliography}
\end{document}